\documentclass[a4paper,11pt]{article}

\usepackage{color}
\usepackage[usenames,dvipsnames]{xcolor}
\usepackage{graphicx}
\usepackage{amsmath}
\usepackage{caption}
\usepackage{siunitx}
\usepackage{booktabs}
\usepackage{tabularx}
\usepackage{multirow}
\usepackage{multicol}
\usepackage{transparent}
\usepackage{placeins}

\usepackage{relsize}

\pdfoutput=1 

\usepackage{jheppub} 

\hypersetup{%
  colorlinks=true, linktocpage=true, pdfstartpage=1, pdfstartview=FitV,
  breaklinks=true, pageanchor=true,
  pdfpagemode=UseNone,
  plainpages=false, bookmarksnumbered, bookmarksopen=true, bookmarksopenlevel=1,
  hypertexnames=true, pdfhighlight=/O,
  urlcolor=RoyalBlue, linkcolor=ForestGreen, citecolor=RoyalBlue,
  pdftitle={Bounds on heavy sterile neutrinos from the angular distribution of B to D* ell nu decays},
  pdfauthor={Florian U. Bernlochner, Marco Fedele, Tim Kretz, Ulrich Nierste, markus T. Prim},
  pdfsubject={},
  pdfkeywords={},
  pdfcreator={pdfLaTeX},
  pdfproducer={LaTeX with hyperref}
}

\newcommand{\mB}{m_B}
\newcommand{\mDs}{m_{D^*}}
\newcommand{\mnu}{m_{\nu}}

\def\cbar{\overline{c}}
\def\sbar{\overline{s}}
\def\lbar{\overline{\ell}}
\def\nubar{\overline{\nu}}
\def\Dst{{D^\ast}}
\def\Bbar{\overline{B}}

\newcommand{\GKstar}[3]{G^{#1, #2}_{#3}}
\newcommand{\GKstart}[3]{\tilde{G}^{#1, #2}_{#3}}
\newcommand{\lambdagast}{\lambda_{\ga^*}}
\newcommand{\Rea}{\textrm{Re}}
\newcommand{\Ima}{\textrm{Im}}
\newcommand{\ga}{\gamma}

\preprint{TTP24-041}

\author[a]{Florian U. Bernlochner,}
\author[b,c]{Marco Fedele,}
\author[b]{Tim Kretz,}
\author[b]{Ulrich Nierste,}
\author[a]{Markus T. Prim}

\affiliation{$^a$ Physikalisches Institut der Rheinischen Friedrich-Wilhelms-Universit\"at Bonn, Nußallee 12, D-53115 Bonn, Germany,\\
$^b$ Institute for Theoretical Particle Physics, Karlsruhe Institute of Technology (KIT),\\
  Wolfgang-Gaede-Str. 1, D-76131 Karlsruhe, Germany\\
$^c$ Departament de Física Teòrica, IFIC, Universitat de València – CSIC,
Parque Científico, Catedrático José Beltrán 2, E-46980 Paterna, Spain}

\emailAdd{florian.urs.bernlochner@cern.ch}
\emailAdd{marco.fedele@ific.uv.es}
\emailAdd{tim.kretz@kit.edu}
\emailAdd{ulrich.nierste@kit.edu}
\emailAdd{markus.prim@cern.ch}

\title{Model independent bounds on heavy sterile neutrinos from the angular distribution of 
$\mathbf{B\to D^*\ell\nu}$ decays}

\abstract{In this paper we study the bounds that can be inferred on New Physics couplings to heavy sterile neutrinos $N$ from the recent measurements performed by the Belle collaboration of the angular analysis of $B\to D^*\ell\bar\nu_\ell$ decays, with $\ell=e,\mu$. Indeed, a sterile neutrino $N$ may lead to competing $B\to D^*\ell\bar N$ decays and Belle might have measured an incoherent sum of these two independent channels. After reviewing the theoretical formalism required to describe this phenomenon in full generality, we first perform a bump hunt in the $M_{\rm miss}^2$ Belle distribution to search for evidences of an additional massive neutrino. We found in such a way a small hint at $M_{\rm miss}^2 \sim (354\ {\rm MeV})^2$. However, the Belle angular analysis is sensitive to $N$ masses up to $\mathcal{O}$(50 MeV), preventing us to further inspect this hint. Nevertheless, we study the potential impact of this additional channel in the allowed mass range on the measured angular distributions and  extract model-independent bounds on the new-physics couplings which could mediate such an interaction. In particular, in the mass window here inspected, we obtain the most stringent bounds for vector and left-handed scalar operators to date.}

\begin{document}

\maketitle
\flushbottom


\section{Introduction}\label{sec:intro}

In the last decade, several measurements of the semileptonic decays Lepton Flavour Universality Violating (LFUV) ratio
\begin{equation}
    R(D^{(*)})= \frac{BR(B\to D^{(*)}\tau\bar\nu)}{BR(B\to D^{(*)}\ell\bar\nu)}\,, \qquad \ell=e,\mu
\end{equation}
have been performed by the BaBar~\cite{BaBar:2012obs,BaBar:2013mob}, Belle~\cite{Belle:2015qfa,Belle:2016ure,Hirose:2016wfn,Hirose:2017dxl, Belle:2019rba}, LHCb~\cite{LHCb:2023zxo,LHCb:2023uiv,LHCb:2024jll} and Belle II~\cite{Belle-II:2024ami} collaborations, resulting in a $\simeq 3 \sigma$ tension~\cite{HFLAV:2022pwe} obtained by HFLAV when confronted with an average of most of the latest SM predictions~\cite{Bigi:2016mdz,Bernlochner:2017jka,Jaiswal:2017rve,Gambino:2019sif,Bordone:2019vic,Martinelli:2021onb}. A recent analysis incorporating novel insights into form factors, even finds a tension of 4.4$\sigma$ with the Standard Model (SM)~\cite{Iguro:2024hyk}. Moreover, the semileptonic $B\to D^{(*)}\ell\bar\nu$ decays with light charged leptons have been among the main decay channels used for the exclusive determination of the $|V_{cb}|$ Cabibbo-Kobayashi-Maskawa (CKM) matrix element. In this context, the most precise determinations of the differential distribution rates come from the Belle~\cite{Belle:2023bwv} and Belle II~\cite{Belle-II:2023okj} collaborations. Even more interestingly, the Belle collaboration recently released, for the first time, the measurement of the full angular differential distribution of $B\to D^{(*)}\ell\bar\nu$ decays with $\ell=e,\mu$~\cite{Belle:2023xgj}.

In order to explain the measured discrepancy in the LFUV ratios, the usually explored approach consists in studying New Physics (NP) extensions of the SM which included the addition of new fields mediating the $b \to c \ell \bar\nu$ transitions, for a recent review see e.g. Ref.~\cite{Capdevila:2023yhq} and references therein. However, another equally interesting NP scenario capable to explain these deviations consists in introducing new states not as mediators, but as actual final states. Indeed, if one extends the SM by the inclusion of Heavy Sterile Neutrinos (HSN), $N$, the observed $b \to c \ell \bar\nu$ decays could actually result as the incoherent sums of two independent channels: on the one hand, the purely SM transition $b \to c \ell \bar\nu_{\ell}$; and on the other hand, a genuine NP decay induced, if kinematically allowed, by the HSN, i.e. $b \to c \ell \bar N$. If the HSN couples (mainly) to taus, this new channel enhances the values of both LFUV ratios, therefore providing a viable explanation of the measured anomalies, as studied in Refs.~\cite{Iguro:2018qzf,Asadi:2018wea,Greljo:2018ogz,Robinson:2018gza,Azatov:2018kzb,Babu:2018vrl,Mandal:2020htr}. 

Nevertheless, even if one is not interested in employing HSN as a means to address the LFUV anomalies, the study of HSN and their interaction with heavy quarks is interesting in its own right. The increasing amount of data on $B\to D^*\ell\nu$ decays can offer rich information regarding HSN extensions of the SM. Indeed, the presence of an HSN will not only have an impact on $R(D^{(*)})$, but also on the increasing amount of information collected from the several differential measurements performed in Refs.~\cite{Belle:2023bwv,Belle-II:2023okj,Belle:2023xgj}.

In this context, the object of this paper is to study the potential impact of HSN on the measurements of the angular $B\to D^*\ell\nu$ distribution, fully performed for the first time in the recent Ref.~\cite{Belle:2023xgj}, with light leptons in the final state. We will carry out our analysis in a model-independent way, i.e., we will be agnostic regarding the UV completion behind the possible origin of such a new particle. To this end, we will employ the following dim-6 effective Hamiltonian below the electroweak scale, where the top quark and the Higgs, $Z$ and $W$ bosons have been integrated out~\cite{Robinson:2018gza}:
\begin{align}
\mathcal{H}_{\rm eff} = \frac{4G_F}{\sqrt2} V_{cb} & \left[ (\cbar_L \gamma_\mu b_L) (\lbar_L \gamma^\mu \nu_{\ell,L}) + 
g_{V_R}^{N,\ell} (\cbar_R \gamma_\mu b_R) (\lbar_R \gamma^\mu N_R)
+ g_{S_L}^{N,\ell} (\cbar_R b_L) (\lbar_L N_R) \right. \nonumber \\
& + g_{S_R}^{N,\ell} (\cbar_L b_R) (\lbar_L N_R)
  + \left. g_{T}^{N,\ell} (\cbar_L \sigma_{\mu\nu} b_R) (\lbar_L \sigma^{\mu\nu} N_R) \right] + \mathrm{h.c.}\,,
\label{eq:Heff_Jp}
\end{align}
where $\psi_{L(R)}\equiv P_{L(R)}\psi$ with $P_{L(R)}\equiv \frac{1\mp\gamma_5}{2}$, we used the convention $\sigma_{\mu\nu}=\frac i2 [\gamma_\mu,\gamma_\nu]$, $G_F$ is the Fermi constant and $V_{cb}$ is the CKM element mediating $b\to c$ transitions. In the effective Hamiltonian above the first operator mediates the SM transition describing the $B\to D^*\ell\nu_\ell$ decay, while the remaining ones are genuine NP operators responsible of the $B\to D^*\ell N$ transition, with $N$ being the HSN.\footnote{Notice that a left-handed vector operator involving a right-handed neutrino is dim-8, therefore not considered in this analysis.} All NP Wilson Coefficients (WCs) are normalized to the SM one and carry a lepton index since we do not want to impose any flavour structure on the NP sector, i.e. LFU is not assumed for HSNs.

In principle, it could be interesting to consider higher order operators which would induce a mixing between the HSN and the SM neutrinos $\nu_e$ and $\nu_\mu$ via the mixing angles $U_{eN}$ and $U_{\mu N}$, respectively. However, below the $B\to D^*\ell N$ kinematical threshold these mixing angles are already strongly constrained, with the largest allowed value found for a HSN with a mass of 2 GeV and corresponding to $|U_{eN}|^2 \simeq |U_{\mu N}|^2 \simeq 10^{-5}$~\cite{Abdullahi:2022jlv}. As we will see below, the angular analysis of the $B\to D^*\ell\nu$ channel has not been measured yet with a precision capable to set competitive bounds. For this reason, we will ignore here the effects coming from the mixing between the HSN and the SM neutrinos.

This paper is organized as follows: in Sec.~\ref{sec:formalism} we will review the formalism employed to describe $b \to c \ell \bar N$ transition, and in Sec.~\ref{sec:Obs} the impact of such a new channel on the measured angular observables will be inspected. Details on the possibility to extract HSN signals compatible with the Belle experimental assumptions are given in Sec.~\ref{sec:Exp}, before bounds on the HSN mass and couplings will be extracted from this channel in Sec.~\ref{sec:bounds}. A comparison with current available flavour bounds is given in Sec.~\ref{sec:comparison}, before conclusions are drawn in Sec.~\ref{sec:conclusions}. Details regarding form factors and the SM helicity amplitudes are relegated to Appendices \ref{app:FF} and \ref{app:SMHA}, respectively.


\section{Theoretical Formalism}\label{sec:formalism}

In this section we will review the theoretical formalism required to describe $B\to D^*\ell\bar N$ decays. To this end, we will adopt the generic formalism developed in Ref.~\cite{Gratrex:2015hna} for any semileptonic meson decay, adapting it to our specific channel. For simplicity, we will report our results below assuming the charged leptons $\ell=e,\mu$ to be massless; nevertheless, the results of our numerical analyses reported in Sec.~\ref{sec:bounds} will take full account of all lepton masses for completeness. Our results match the ones previously obtained in Ref.~\cite{Datta:2022czw}.


\subsection{The Helicity Amplitudes}\label{sec:Hamp}

As a first step, we will introduce here the Helicity Amplitudes (HAs). These objects are defined in terms of 7 form factors ($V$, $A_0$, $A_1$, $A_{12}$, $T_1$, $T_2$, $T_{23}$), whose details regarding their definition and determination are given in Appendix \ref{app:FF}. For the sake of brevity, we omit the $q^2$ dependence from all form factors here and below. $q^2$ is the square of the invariant mass of the $(\ell,N)$ pair. Moreover we stress that, while we are not explicitly writing it to keep the notation simpler, these HAs are not lepton flavour universal due to our choice to not impose any flavour structure in the NP sector. Under those assumptions, the HAs read
\begin{align}
H^{N}_{V,0} &= -i \frac{2 \mB \mDs }{\sqrt{q^2}} A_{12} \; g_{V_R}^{N,\ell} \,,  \nonumber\\ 
H^{N}_{V,\pm} & =\frac{i}{4 \left(\mB + \mDs\right)} \left(\pm \sqrt{\lambda_B} \, V + \left(\mB + \mDs\right)^2 A_1 \right) g_{V_R}^{N,\ell} \,,   \nonumber  \\
H_P^{N} &= -\frac{i \sqrt{\lambda_B}}{4} A_0 \left( \frac{ g_{S_L}^{N,\ell} - g_{S_R}^{N,\ell}}{m_b + m_c} -
\frac{m_N}{q^2} g_{V_R}^{N,\ell} \right) \,, \nonumber \\
H_S^{N} &= \frac{i \sqrt{\lambda_B}}{4} A_0 \left( \frac{ g_{S_L}^{N,\ell} - g_{S_R}^{N,\ell}}{m_b + m_c} + 
\frac{m_N}{q^2} g_{V_R}^{N,\ell} \right) \,, \nonumber
\end{align}
\begin{align}
\label{eq:HHAs}
H^{N}_{T,0} & = \frac{ 2 \sqrt{2} \mB \mDs}{\mB + \mDs} T_{23} \; g_{T}^{N,\ell} \,,  \nonumber\\
H^{N}_{T,\pm} &= \frac{1}{\sqrt{2 q^2} } \left(\mp \sqrt{ \lambda_B} T_1 -  \left( \mB^2 - \mDs^2 \right) T_2 \right) g_{T}^{N,\ell} \,.
\end{align}
In the above equations, $m_N$ is the HSN mass, $m_B$ and $m_{D^*}$ are the masses of the $B$ and the $D^*$ mesons, respectively, and $m_b$ and $m_c$ are the masses of the bottom and charm quarks. Defining the K\"all\'en-function as $\lambda(a,b,c) = a^2 + b^2 + c^2 - 2(ab+ac+bc)$, above we employed the quantity $\lambda_B\equiv \lambda(q^2, \mB^2, \mDs^2)$.

As it can be seen when confronting with Appendix \ref{app:SMHA}, where the SM HAs are listed, the presence of tensor HAs $H^{T,Tt}_\lambda$, with $\lambda=0,\pm$, is a genuine NP effect induced by the presence of HSNs. Something similar can be said about the (pseudo)scalar ones $H_{P,S}$, which in the SM are proportional to the charged lepton mass and hence negligible for $\ell=e,\mu$. Therefore, when studying potential effects coming from HSNs, the knowledge of all 7 form factors is required, contrarily to the SM case where only 3 (4) form factors are necessary in order to describe $B\to D^*\ell\bar\nu$ decays with massless (massive) charged leptons.


\subsection{The Angular Coefficients}\label{sec:Acoeffs}

We have now all the ingredients required to build the Angular Coefficients (ACs) $G^{lk}_m$, where $l$ is a non-negative integer index running up to twice the $D^*$ spin, i.e. $0 \leq l \leq 2$, $k$ is a non-negative integer index running up to twice the $(\ell\bar\nu)$ system spin, i.e. $0 \leq l \leq 2$, and $m$ is the coherent sum of the helicities of all internal particles. Notice that, due to the final meson states $D$ and $\pi$ being pseudoscalars, ACs with $l=1$ are forbidden.

Before giving the explicit expressions for ACs in terms of the HAs, it is useful to introduce the following notations:
\begin{equation}
E^\nu_1 = \frac{-\mnu^2+q^2}{2\sqrt{q^2}}\,,\qquad E^\nu_2 = \frac{\mnu^2+q^2}{2\sqrt{q^2}}\,, \qquad \lambdagast^\nu\equiv\lambda(q^2,\mnu^2,0)\,,
\end{equation}
where $\nu$ is either the massless SM neutrino $\nu_\ell$, for which $m_{\nu_\ell}=0$, or a HSN $N$, in which case $m_N\neq 0$. We also observe that all the ACs share the same normalization $\mathcal{N}$, such that it is possible to define their normalized version $ \tilde{G}^{lk}_m$ as
\begin{equation}
    G^{lk}_m \equiv \mathcal{N} \tilde{G}^{lk}_m\,, \qquad \text{with} \qquad \mathcal{N} = \frac43\left( \frac{4G_F}{\sqrt{2}}V_{cb} \right)^2 \frac{\sqrt{\lambda_B \lambdagast^\nu}}{2^9\pi^3\mB^3q^2}\,.
\end{equation}
With the above notation, and stressing once again that also the ACs are lepton flavour specific due to their dependence on HAs, i.e. $G^{lk}_m(e) \neq G^{lk}_m(\mu)$, the normalized $\tilde{G}^{lk}_m$ for a generic lepton can be written as
\begin{alignat}{2}
\label{eq:explicitGdifferentmass}
&\tilde{G}^{00}_0 = \sum_{\nu} &&\frac{2}{3}\left( 3 E_1^\nu E_2^\nu+ \frac{\lambdagast^\nu}{4q^2}\right) \left(\left|H^{\nu}_{V,+}\right|^2+\left|H^{\nu}_{V,-}\right|^2 + \left|H^{\nu}_{V,0}\right|^2 \right)  \nonumber \\
& && + \left(  E_1^\nu E_2^\nu +\frac{\lambdagast^\nu}{4q^2} \right) \left( \left|H_S^\nu\right|^2 +  \left|H_P^\nu\right|^2 \right) \nonumber \\
& && + \frac{4}{3} \left( 3 E_1^\nu E_2^\nu - \frac{\lambdagast^\nu}{4q^2} \right) \left(  \left|H_{T,+}^{\nu}\right|^2+\left|H_{T,-}^{\nu}\right|^2+\left|H_{T,0}^{\nu}\right|^2\right) \,, \\[0.2cm]\label{eq:G000}
&\GKstart{0}{1}{0} = \sum_{\nu} &&\sqrt{\lambdagast^\nu} \Bigg( \left|H^{\nu}_{V,-}\right|^2 - \left|H^{\nu}_{V,+}\right|^2 - 2 \frac{\mnu^2}{q^2} \left(\left|H_{T,-}^{\nu}\right|^2 - \left|H_{T,+}^{\nu}\right|^2\right) \nonumber \\ 
& && - \frac{\mnu}{\sqrt{q^2}} \Rea \left[ H_{V,0}^{\nu} (\bar H_P^\nu + \bar H_S^\nu) \right]  + \sqrt{2}  \Ima \left[H_{T,0}^{\nu} (\bar H_P^\nu - \bar H_S^\nu)\right]  \Bigg) \,, \\[0.2cm]
&\GKstart{0}{2}{0} = \sum_{\nu} &&-\frac{1}{3} \frac{\lambdagast^\nu}{q^2} \Bigg(2 \left|H_{V,0}^{\nu} \right|^2 - \left|H_{V,+}^{\nu} \right|^2 - \left|H_{V,-}^{\nu} \right|^2 - 2 \left(2 \left|H_{T,0}^{\nu}\right|^2 - \left|H_{T,+}^{\nu}\right|^2- \left|H_{T,-}^{\nu}\right|^2 \right)  \Bigg) \;, \displaybreak[0] \\[0.2cm]
&\GKstart{2}{0}{0} = \sum_{\nu} &&-\frac{2}{3}\left( 3 E_1^\nu E_2^\nu+ \frac{\lambdagast^\nu}{4q^2}\right) \left(\left|H_{V,+}^{\nu}\right|^2+\left|H_{V,-}^{\nu}\right|^2 - 2\left|H_{V,0}^{\nu}\right|^2 \right)  \nonumber \\
& && + 2 \left(  E_1^\nu E_2^\nu  + \frac{\lambdagast^\nu}{4q^2} \right) \left( \left|H_S^\nu\right|^2 +  \left|H_P^\nu\right|^2 \right) \nonumber \\
& && - \frac{4}{3} \left( 3 E_1^\nu E_2^\nu - \frac{\lambdagast^\nu}{4q^2} \right) \left(  \left|H_{T,+}^{\nu}\right|^2+\left|H_{T,-}^{\nu}\right|^2 - 2 \left|H_{T,0}^{\nu}\right|^2\right) \; , \displaybreak[0] \\[0.2cm]
&\GKstart{2}{1}{0} = \sum_{\nu} &&-\sqrt{\lambdagast^\nu} \Bigg( \left|H^{\nu}_{V,-}\right|^2 - \left|H^{\nu}_{V,+}\right|^2 - 2 \frac{\mnu^2}{q^2} \left(\left|H_{T,-}^{\nu}\right|^2 - \left|H_{T,+}^{\nu}\right|^2\right) \nonumber \\ 
& && + 2\frac{\mnu}{\sqrt{q^2}} \Rea \left[ H_{V,0}^{\nu} (\bar H_P^\nu + \bar H_S^\nu) \right]  - 2 \sqrt{2} \, \Ima \left[H_{T,0}^{\nu} (\bar H_P^\nu - \bar H_S^\nu) \right]  \Bigg) \; , \displaybreak[0] \\[0.2cm]
&\GKstart{2}{2}{0} = \sum_{\nu} && -\frac{1}{3} \frac{\lambdagast^\nu}{ q^2 } \Bigg(4 \left|H_{V,0}^{\nu} \right|^2 + \left|H_{V,+}^{\nu} \right|^2 + \left|H_{V,-}^{\nu} \right|^2 - 2 \left(4 \left|H_{T,0}^{\nu}\right|^2 + \left|H_{T,+}^{\nu}\right|^2 + \left|H_{T,-}^{\nu}\right|^2 \right)  \Bigg) \;, \displaybreak[0]  \\[0.2cm]
&\GKstart{2}{1}{1} = \sum_{\nu} && \sqrt{3\lambdagast^\nu} \Bigg( 2 \left( H_{V,0}^{\nu} \bar H_{V,-}^{\nu} - H_{V,+}^{\nu} \bar H_{V,0}^{\nu} \right) + i \sqrt{2} \left( H_{T,+}^{\nu} (\bar H_P^\nu - \bar H_S^\nu) - (H_P^\nu - H_S^\nu) \bar H_{T,-}^{\nu} \right) \nonumber \\
& && + \frac{\mnu}{\sqrt{q^2} } \left( H_{V,+}^{\nu} (\bar H_P^\nu + \bar H_S^\nu) + (H_P^\nu + H_S^\nu) \bar H_{V,-}^{\nu} \right) + 4 \frac{\mnu^2}{q^2} \left( H_{T,+}^{\nu} \bar H_{T,0}^{\nu} - H_{T,0}^{\nu} \bar H_{T,-}^{\nu} \right) \Bigg)  \; , \displaybreak[0] \label{eq:G211}\\[0.2cm]
&\GKstart{2}{2}{1}= \sum_{\nu} && 2 \frac{\lambdagast^\nu}{ q^2}  \Bigg( H_{V,+}^{\nu} \bar H_{V,0}^{\nu} + H_{V,0}^{\nu} \bar H_{V,-}^{\nu} - 2 \left( H_{T,+}^{\nu} \bar H_{T,0}^{\nu}+H_0^{T,\nu} \bar H_-^{T,\nu} \right)\Bigg) \; , \displaybreak[0] \label{eq:G221}\\[0.2cm]
& \GKstart{2}{2}{2} = \sum_{\nu} && - 4 \frac{\lambdagast^\nu}{ q^2} \Bigg(H_{V,+}^{\nu} \bar H_{V,-}^{\nu}  -2 H_{T,+}^{\nu} \bar H_{T,-}^{\nu} \Bigg) \; . \displaybreak[0]\label{eq:G222}
\end{alignat}
In the above Equations, we intend the sum over $\nu$ as the sum over the SM HAs, given in Appendix \ref{app:SMHA}, and NP ones, reported in Sec.~\ref{sec:Hamp}.


\section{The impact of heavy sterile neutrinos on the angular observables}\label{sec:Obs}

Combining all the ingredients introduced in Sec.~\ref{sec:formalism}, we can now write the fully differential distribution as
\begin{alignat}{1}
\label{eq:d4GJi}
\frac{32 \pi}{9} \frac{d^4 \Gamma}{dq^2\,d\textrm{cos}\theta_{\ell} \, d\textrm{cos}\theta_V \, d \phi }  = & \left(J_{1s} + J_{2s}\cos 2\theta_{\ell} + J_{6s} \cos \theta_{\ell} \right) \sin^2 \theta_V \,+  \nonumber \\
&{} \left(J_{1c} + J_{2c}\cos 2\theta_{\ell} + J_{6c} \cos \theta_{\ell} \right) \cos^2 \theta_V \,+  \nonumber \\
&{} \left(J_3 \cos 2 \phi +J_9  \sin 2\phi\right)\sin^2 \theta_V \sin^2 \theta_{\ell} \, +  \nonumber \\
&{} \left(J_4 \cos  \phi +J_8  \sin \phi\right) \sin 2 \theta_V \sin 2\theta_{\ell}  \,+  \nonumber \\
&{} \left(J_5 \cos  \phi +J_7  \sin \phi\right) \sin 2 \theta_V \sin \theta_{\ell} \, , 
\end{alignat}
where $\theta_{\ell}$ is the angle between the charged lepton and the direction opposite the $B$ meson in the $W^{*}$ boson rest frame, $\theta_V$ is the angle between the direction opposite the $B$ meson and the $D$ meson in the $D^*$ rest frame, and $\phi$ is the angle between the planes spanned by the $(W^*,\ell)$ and $(D,D^*)$ systems in the $B$ meson rest frame. Above, the $J_i$ ACs are related to the ones introduced in the previous section through the relations
\begin{alignat}{2}
\label{eq:AngObs}
&J_{1s} = \frac{\Rea(8\GKstar{0}{0}{0} + 2\GKstar{0}{2}{0} - 4 \GKstar{2}{0}{0} - \GKstar{2}{2}{0})}{3}   \,, \qquad \qquad 
&&J_{2s} = \Rea(2\GKstar{0}{2}{0} - \GKstar{2}{2}{0} ) \,, \nonumber\\
&J_{1c} = \frac{\Rea(8\GKstar{0}{0}{0} + 2\GKstar{0}{2}{0} + 8 \GKstar{2}{0}{0} + 2 \GKstar{2}{2}{0})}{3} \,, \qquad 
&&J_{2c} = 2\,\Rea(\GKstar{0}{2}{0} + \GKstar{2}{2}{0} )  \,, \nonumber\\
&J_{6s} = - \frac{\Rea(8\GKstar{0}{1}{0} - 4\GKstar{2}{1}{0})}{3} \,, 
&&J_{6c} = - \frac{8\, \Rea(\GKstar{0}{1}{0} + \GKstar{2}{1}{0})}{3} \,, \nonumber\\
&J_{3} = \Rea(\GKstar{2}{2}{2}) \,, \qquad \qquad \quad 
J_{4} = -\Rea(\GKstar{2}{2}{1})  \,, \quad 
&&J_{5} = \frac{2}{\sqrt{3}}\Rea(\GKstar{2}{1}{1}) \,, \nonumber\\
&J_{9} = -\Ima(\GKstar{2}{2}{2}) \,, \,\qquad \quad \quad  
J_{8} = \Ima(\GKstar{2}{2}{1})  \,, \quad 
&&J_{7} = -\frac{2}{\sqrt{3}}\Ima(\GKstar{2}{1}{1}) \,.
\end{alignat}
Given those definitions, it is possible to define the branching fraction $d\Gamma/dq^2$, the forward-backward asymmetry $A_{\rm FB}$ and the longitudinal polarisation fraction $F_L$ as:
\begin{alignat}{2}
\frac{d\Gamma}{dq^2} &= 6\,  \GKstar{0}{0}{0} =  \frac{3 J_{1c} + 6 J_{1s} - J_{2c} - 2 J_{2s}}{4} \,, \\
A_{\rm FB} &= \frac{1}{2} \frac{ \GKstar{0}{1}{0} }{\GKstar{0}{0}{0} } = -\frac{3}{8} \frac{ J_{6c} + 2 J_{6s} }{d\Gamma/dq^2} \,, \\
F_L &= \frac{\GKstar{0}{0}{0}+\GKstar{2}{0}{0}}{3\GKstar{0}{0}{0}} = \frac{3 J_{1c} - J_{2c}}{4\,d\Gamma/dq^2}\,.
\end{alignat}
The Belle collaboration~\cite{Belle:2023xgj} measured ACs in different bins of $w(q^2) = (m_{B}^2+m_{D^{*}}^2 - q^2)/(2m_{B}m_{D^{*}})$, reporting integrated values for the coefficients over different bin ranges, namely
\begin{equation}
\bar J_i^{(n)} = \int_{\Delta w^{(n)}} dw J_i(w)\,,
\end{equation}
where $\Delta w^{(n)}$ are the ranges of the $n$-th bin. Moreover, the quantities actually determined are normalized to the total decay rate according to the following relation:
\begin{equation}\label{eq:Jhat}
\hat J_i^{(n)} = \frac{\bar J_i^{(n)}}{\int_{w_{\rm min}}^{w_{\rm max}} dw\; d\Gamma/dw}\,.
\end{equation}

A few considerations are now in order. As already stated before, the Belle analysis is sensitive to potential contributions from HSNs only if the $B\to D^*\ell N$ decay is kinematically allowed. This implies that the heaviest mass that can in principle be probed via this analysis (assuming massless charged leptons) corresponds to $m_N^{\rm max}= (m_B-m_{D^*})^2 \simeq 3$ GeV. However, this is actually not the case due to experimental limitations in the way the angular analysis has been performed at Belle. Further details can be found in Sec.~\ref{sec:Exp}.

Nevertheless, it is interesting to observe that in general the effect of HSNs are not appreciable only in bins whose range is above the HSN mass, even if its effects are kinematically forbidden for values of $q^2 < m_N^2$. Indeed, due to the fact that Belle measured the binned ACs normalized to the full decay rate as shown in Eq.~\eqref{eq:Jhat}, all measured coefficients ratios will be sensitive to the HSN effect.

Moreover, the presence of HSNs has interesting consequences on the relations among the first 4 ACs. Indeed, combining Eq.~\eqref{eq:AngObs} with Eqs.~\eqref{eq:G000}-\eqref{eq:G222} and taking the $m_\nu\to 0$ limit, it is possible to write
\begin{align}
J_{1s}^\nu &= q^2 \left( 3 \left|H_{V,+}^{\nu}\right|^2 + 3\left|H_{V,-}^{\nu}\right|^2 + 2 \left|H_{T,+}^{\nu}\right|^2 + 2\left|H_{T,-}^{\nu}\right|^2 \right) \,,\\
J_{2s}^\nu &= q^2 \left( \left|H_{V,+}^{\nu}\right|^2 + \left|H_{V,-}^{\nu}\right|^2 - 2 \left|H_{T,+}^{\nu}\right|^2 - 2\left|H_{T,-}^{\nu}\right|^2 \right) \,,\\
J_{1c}^\nu &= 4 q^2 \left( \left|H_{V,0}^{\nu}\right|^2 + 2 \left|H_{T,0}^{\nu}\right|^2 + \left|H_{S}^{\nu}\right|^2 + \left|H_{P}^{\nu}\right|^2 \right) \,,\\
J_{2c}^\nu &= 4 q^2 \left( - \left|H_{V,0}^{\nu}\right|^2 + 2 \left|H_{T,0}^{\nu}\right|^2 \right)\,.
\end{align}
In the SM the above equations further simplify due to the tensorial HA being vanishing and the (pseudo)scalar ones being proportional to the lepton mass and hence negligible, inducing the following relations:
\begin{equation}\label{eq:J_rel}
J_{1s}^{\nu_\ell} = 3J_{2s}^{\nu_\ell}\,, \qquad \qquad \qquad J_{1c}^{\nu_\ell} = -J_{2c}^{\nu_\ell}\,.
\end{equation}
These relations might however be violated once an HSN is introduced in the theory. Indeed, even in the case of an HSN with a mass similar to the muon one, the above relations hold only if the interaction between the HSN and the SM is mediated by the $g_{V_R}^{N,\ell}$ coupling, which produces contributions to the HAs analogous to the SM ones, cf. Eqs.~\eqref{eq:HHAs} and \eqref{eq:HHAs_SM}. Conversely, the presence of any of the other couplings would violate the above relations: on the one hand, introducing the $g_{S_L(R)}^{N,\ell}$ coupling would lift the helicity suppression in the (pseudo)scalar HAs hence violating the second relation given at Eq.~\eqref{eq:J_rel}; on the other hand, the introduction of the $g_{T}^{N,\ell}$ coupling would violate both the equalities of Eq.~\eqref{eq:J_rel}. Testing the validity of Eq.~\eqref{eq:J_rel} would therefore correspond to testing the presence of HSNs interacting with the SM via a (pseudo)scalar or tensorial current.

Finally, it is worth to observe that the fate of the observables $\hat J_{7,8,9}$, which are sensitive to imaginary parts of the WCs as shown in Eq.~\eqref{eq:AngObs} and hence null tests in the SM, is similar for the case of HSNs. Starting from $\hat J_{8}$ and $\hat J_{9}$, it is straightforward to infer from Eqs.~\eqref{eq:G221}-\eqref{eq:G222} that those ACs are proportional to absolute values of NP WCs: we therefore obtain that $\hat J_{8}$ and $\hat J_{9}$ are null tests for this specific extension of the SM as well. Concerning $\hat J_{7}$,  it is evident from Eq.~\eqref{eq:G211} that a non-vanishing value for this observable is allowed only in the presence of an interference within either a vector or a tensor HA, and a (pseudo)scalar one. This means that a non-vanishing value for this observable is allowed only in the presence of multiple non-vanishing WCs, while in an analysis performed allowing only one WC at a time this observable will also remain equal to zero.


\section{Sterile Neutrinos in the Belle data}\label{sec:Exp}

\begin{figure}
    \centering
    \includegraphics[width=1.0\linewidth]{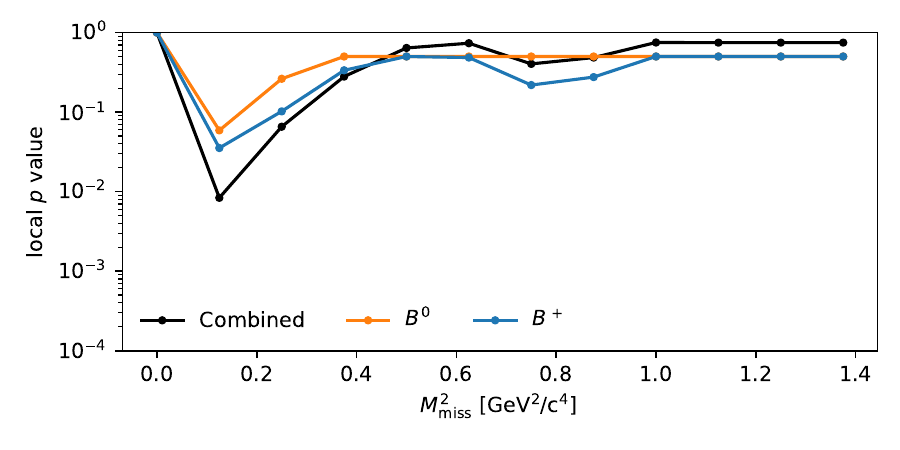}
    \caption{The extracted local $p$-value for a sterile neutrino signal based on the $M_\mathrm{miss}^2$ distribution from Ref.~\cite{Belle:2023bwv}. The local $p$-value is determined every $0.125\,\mathrm{GeV}^2/c^4$, indicated by the points on the plot. The pacing corresponds to the bin width in the available $M_\mathrm{miss}^2$ distribution, which is averaged over $\ell=e,\mu$.}
    \label{fig:mm2limit}
\end{figure}

\begin{figure}
    \centering
    \includegraphics[width=1.0\linewidth]{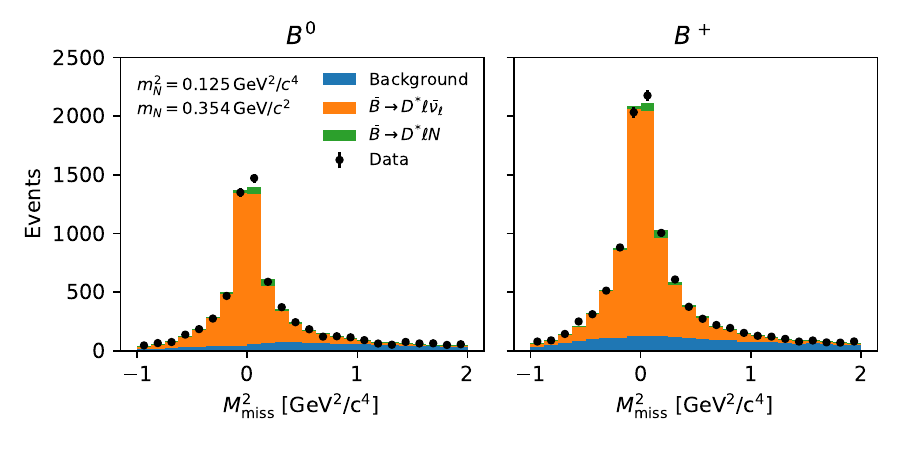}
    \caption{Our fit of the $M_\mathrm{miss}^2$ distribution from Ref.~\cite{Belle:2023bwv}, where the $\bar{B}\to D^*\ell \bar{\nu}_\ell$ components are aggregated into the orange template and all backgrounds are aggregated into the blue template. The green template shows the injected $\bar{B}\to D^*\ell N$ template with $m_N=0.354\,\mathrm{GeV/c^2}$. This fit corresponds to the most significant local $p$-value found in our scan of the $M_\mathrm{miss}^2$ distribution.}
    \label{fig:mm2fit}
\end{figure}

The existence of the decay $B\to D^*\ell N$ would have an implication on the measured missing mass squared distribution $M_\mathrm{miss}^2$~\cite{Kim:2019xqj}. To this end, we investigate the published $M_\mathrm{miss}^2$ distribution by Belle, which is averaged over $\ell=e,\mu$~\cite{Belle:2023bwv}. This distribution is reconstructed using hadronic tagging, and at the $B$-factories it corresponds to the squared mass of the neutrino in the $B\to D^*\ell \nu$ decay. In the presence of a $B\to D^*\ell N$ decay, a second peak would emerge at the position of the HSN mass $m_N^2$; we therefore digitized the $M_\mathrm{miss}^2$ spectrum in order to look for this second peak via a scan. In doing so, we assume that the shape of the HSN peak is identical to the shape of the Standard Model neutrino peak at $M_\mathrm{miss}^2=0$. For our scan, we shift the HSN template starting from zero with a step size corresponding to the bin width in which the data is provided. We perform the scan both independently for the $B^0$ and $B^+$ data and for the combination of the two, where we do not correlate any of the parameters between the datasets. We only allow for positive $B\to D^*\ell N$ signal yields and use the test statistic $q_0 = -2\ln \mathcal{L}_\mathrm{NP} + 2\ln \mathcal{L}_\mathrm{SM}$, which is asymptotically distributed as
\begin{equation}
\begin{aligned}
    f(q_0) &= \frac{1}{2} \delta(q^0) + \frac{1}{2} \chi^2(q^0; 1\,\mathrm{dof}) \,,\\
    f(q_0) &= \frac{1}{4} \delta(q^0) + \frac{1}{2} \chi^2(q^0; 1\,\mathrm{dof}) + \frac{1}{4} \chi^2(q^0; 2\,\mathrm{dof})\,,
\end{aligned}
\end{equation}
for the individual fits and for the combined fits, respectively. See Ref.~\cite{brazzale2024likelihood} for further details about asymptotic distributions. The result of the scan is shown in Fig.~\ref{fig:mm2limit}, while the fit in $M_\mathrm{miss}^2$ is shown in Fig.~\ref{fig:mm2fit} for the HSN mass hypothesis with the largest significance.


Due to Belle not providing the absolute normalization for the MC distribution, we can only perform a bump hunt and quote the local $p$ value, and not transform this into an upper limit for the $\bar{B} \to D^* \ell \bar{\nu}_\ell$ branching ratio. However, we can draw three conclusions: first, at high $M_\mathrm{miss}^2$ values there is almost no room for a sterile neutrino. Second, at $M_\mathrm{miss}^2\approx 0$ the SM neutrino and the sterile neutrino cannot be disentangled in $M_\mathrm{miss}^2$. This is due to the fact that these two templates become indistinguishable as $m_N \to 0$. Third, at small $M_\mathrm{miss}^2\sim (354\,\mbox{MeV})^2$  we see a preference for a sterile neutrino contribution over a pure SM process.

To increase sensitivity to a sterile neutrino final state we can investigate, for the heavy neutrino mass range $m_N \in $ [0--62.5] MeV,\footnote{Heavier sterile neutrino masses would result already in a shift of the $p$-value distribution shown in Figure 5 in Ref.~\cite{Belle:2023bwv} from uniform to peaking at 0, and not to the expected change in the signal yield one would naively expect. This is caused because the signal extraction in the Belle analysis has no sterile neutrino template included and would result in significant biases in the extracted signal yields if the sterile neutrino would be present in the data, and therefore make any interpretation with large $m_N$ unreliable. The region at small $M_\mathrm{miss}^2$ is however ideally probed with the angular coefficients, as for $m_N \to 0$ the sterile and the SM neutrino cannot be discriminated. This is discussed in the main text.} the angular coefficients measured by the Belle collaboration~\cite{Belle:2023xgj}. Indeed, the Belle signal yield extraction in this mass range could be biased up to $\approx 12\%$ and would further increase when higher sterile neutrino masses are used, which would lead to biased angular coefficients. We decided a bias of $\approx 12\%$ is acceptable for the study performed in the main text, but the actual upper bound on the interval is chosen somewhat arbitrarily and should be kept in mind when interpreting the results.


\section{The New-Physics bounds}\label{sec:bounds}

We will now report the results of our NP fits to the full differential angular data first measured for light leptons by the Belle collaboration in Ref.~\cite{Belle:2023xgj}. In more details, we included in our fits data concerning the electron and muon channel separately, neglecting isospin breaking effects due to current level of precision. Considering that the 12 angular observables have been measured in 4 different $w$ bins each, this means that we are in principle fitting for 48 observables in the $B\to D^*e\bar\nu$ channel and 48 observables in the $B\to D^*\mu\bar\nu$ one, with $B\to D^*$ being an average of $\bar{B}^0\to D^{*+}$ and $B^-\to D^{*0}$. However, remembering from Eq.~\eqref{eq:Jhat} that the quantities actually measured are normalized to the total decay rate, which is function of 4 of the 12 angular coefficients, in the fit we have to remove one observable from each channel in order to avoid double-counting. We chose, without any loss of generality, this observable to be $\hat J_{2c}^{(3)}$. In conclusion, we performed fits to a total of 94 angular observables, with the full correlation matrix taken into account as provided in the HEPdata associated to Ref.~\cite{Belle:2023xgj}.

Our analyses are performed in the Bayesian framework employing Markov Chain Monte Carlo fits. This task has been carried out by implementing the analytic expressions given in Sec.~\ref{sec:formalism} in the \texttt{HEPfit} code~\cite{DeBlas:2019ehy}, with full dependence on charged lepton masses taken into consideration for completeness. The SM parameters varied in the fits are the ones concerning the description of the form factors, for which we employed the correlated lattice results estimated by the JLQCD collaboration~\cite{Aoki:2023qpa}, see Appendix~\ref{app:FF} for further details. Concerning each NP scenario here investigated, we assumed flat priors for the their specific parameters. Following the discussion in Sec.~\ref{sec:Exp}, the range allowed for the heavy neutrino mass $m_N$ is [0--62.5] MeV. Contrarily, for the NP WCs parameterized as $g_i\equiv |g_i|e^{i\phi_i}$, we allowed the priors of their absolute values to be large enough, in such a way that their posterior distributions would neither be cut (given the absence of any motivated reason to do so, in contrast to the $m_N$ case), nor would be modified by further enlarging such ranges. Concerning their phases, we allowed them to be flat in the full [-$\pi$, $\pi$] range. 

Finally, we also performed a model comparison between the SM scenario and the several NP ones employing the information criterion~\cite{IC}, defined as:
\begin{equation}\label{eq:IC}
   IC \equiv -2 \overline{\log \mathcal{L}} \, + \, 4 \sigma^{2}_{\log \mathcal{L}} \,,
\end{equation}
where the first and second terms are the mean and variance of the log-likelihood posterior distribution, respectively. The former term measures the quality of the fit, with the latter one being a penalty factor counting effectively the number of the model parameters, therefore penalizing more complicated models. While the overall normalization of the IC is an unknown quantity depending on the experimental data included in the likelihood, Eq.~\eqref{eq:IC} implies that, when fitting to the same dataset, models with smaller IC should be preferred over models with higher ones: it is therefore customary to perform model comparisons by looking at IC differences, namely $\Delta IC\equiv IC_{\rm NP} - IC_{\rm SM}$, with a positive (negative) value for $\Delta IC$ implying a preference for the SM (NP) scenario~\cite{BayesFactors}.

In the rest of this Section we will illustrate the results of our fits. As we have assumed the NP not to be LFU, we will discuss separately the scenarios where the HSNs are produced in association with an electron or a muon, respectively. For each case we will inspect 5 different scenarios, differentiated by whether we allow for NP effects to be present in only one of the 4 different operators listed in Eq.~\eqref{eq:Heff_Jp} at a time, or rather allowing for all of them at the same time.


\subsection{New Physics in the electron channel}\label{sec:NP_e}

\begin{figure*}[!t!]
\centering
\includegraphics[width = 0.45\textwidth]{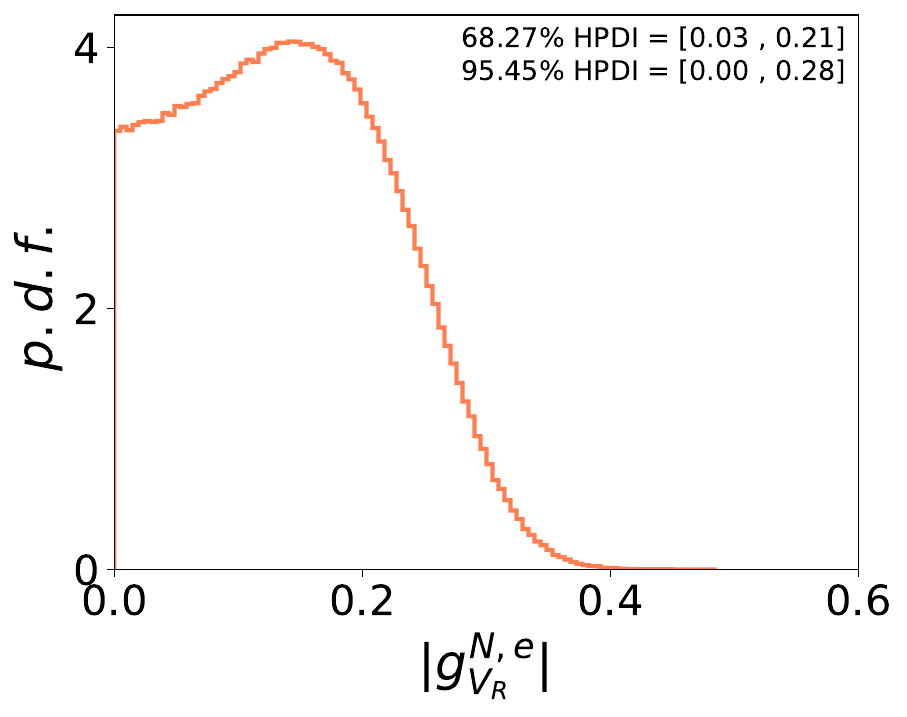}\hspace{1.4em}
\includegraphics[width = 0.47\textwidth]{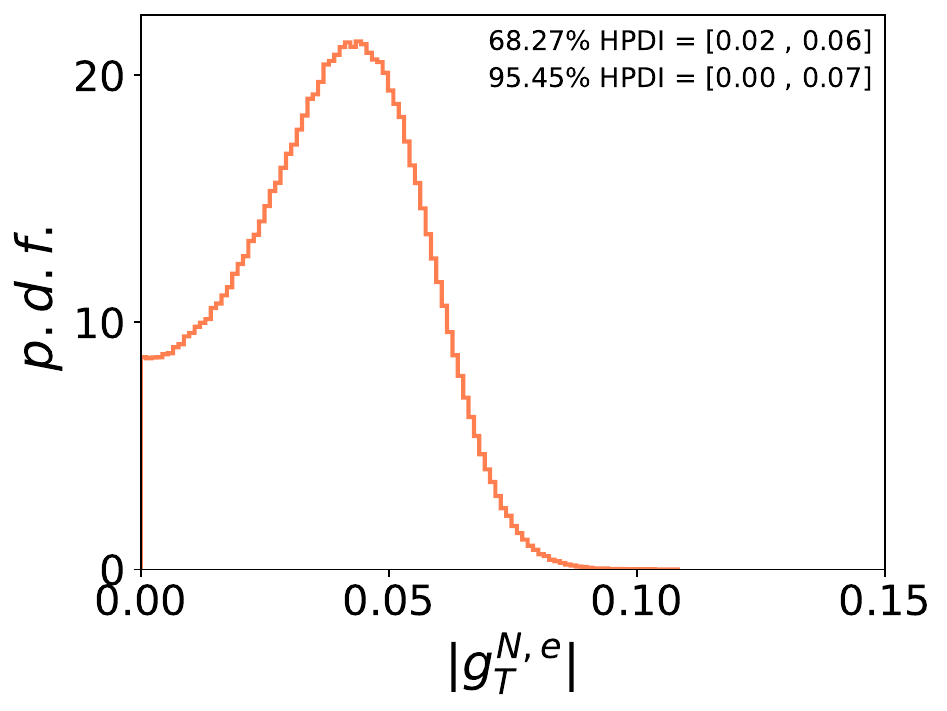}\\
\includegraphics[width = 0.45\textwidth]{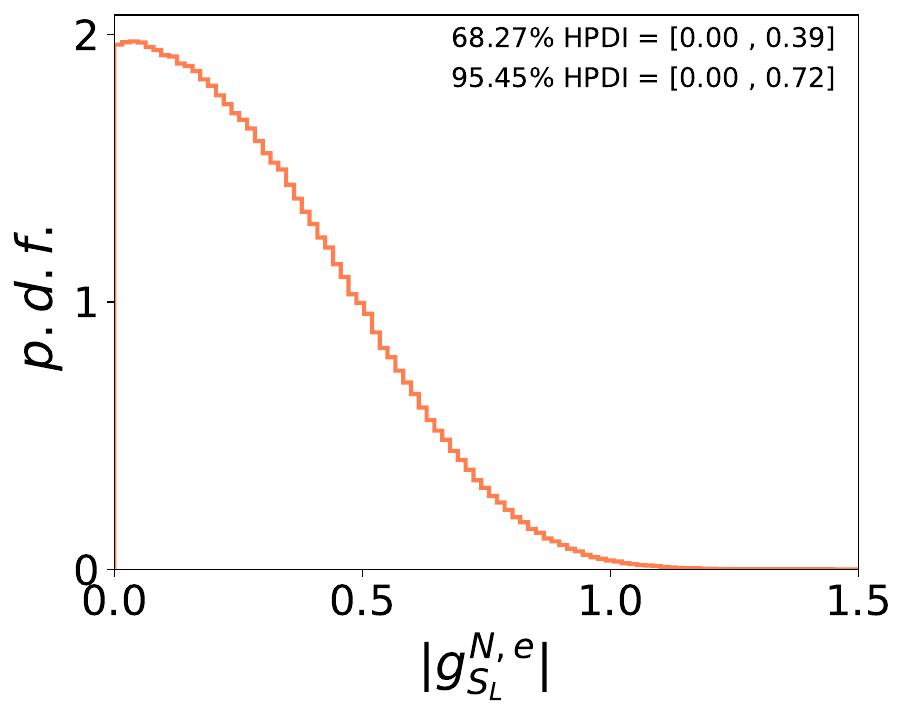}\hspace{2em}
\includegraphics[width = 0.45\textwidth]{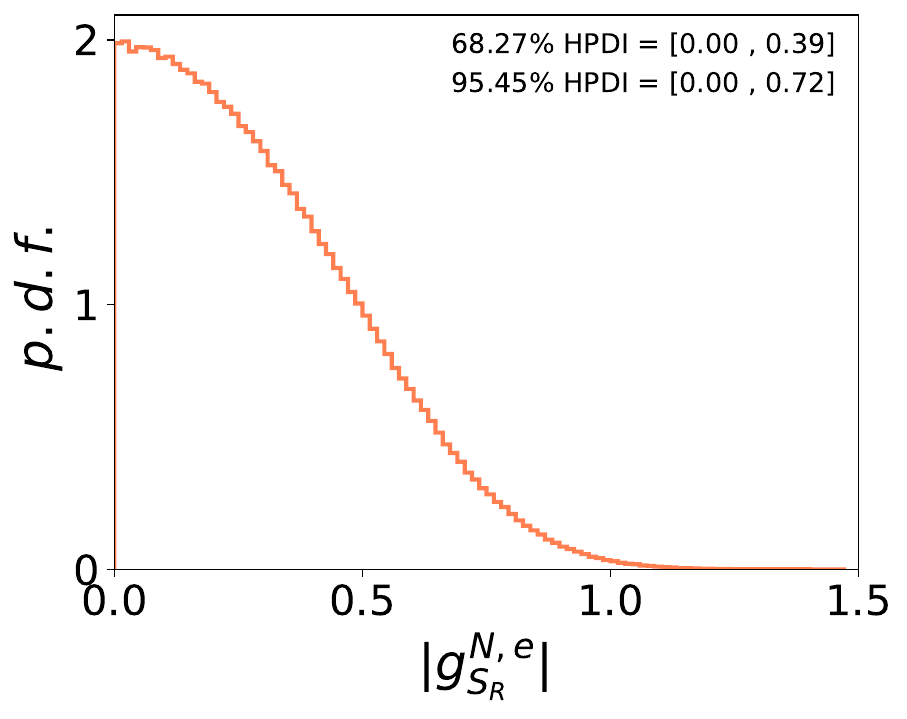}
\caption{Normalised posterior distribution functions for the absolute values of NP WCs where the HSN is produced in association with an electron. These distributions are obtained separately, allowing for one non-vanishing WC at a time respectively. 68.27\% and 95.45\% HPDI are reported as well, see the text for further details.}
\label{fig:1WC_e}
\end{figure*}

As a first step of our study, we focus on the case where HSN are produced in association with electrons, starting from scenarios where only one WC at a time is varied in the fit. Remembering the discussion at the end of Sec.~\ref{sec:Obs}, we set in these cases $\phi_i=0$, given the insensitivity to NP phases in the single WC scenarios. As a first result, we observe that no preference for a particular value of the HSN mass has been found, with the \emph{probability distribution function (p.d.f.)} describing its posterior being flat over the whole scanned region in each of the studied scenarios. Moreover, no meaningful correlation between the HSN mass and any of the WCs absolute values was observed. Concerning the latter, fit results relative to each of the investigated scenarios are reported in Fig.~\ref{fig:1WC_e}, where the marginalised \emph{p.d.f.s} for the WCs absolute values are shown, together with their 68.27\% and 95.45\% \emph{highest posterior density interval} (HPDI). As it can be seen from each of the 4 panels, in none of these scenarios an evidence for NP was observed. This is due to fact that the angular data measured by Belle observed no real deviation from their SM predictions. Such a conclusion is also corroborated by the values of the $\Delta IC$ obtained for the various scenarios, which are reported in Tab.~\ref{tab:DIC_e} and show a preference of the SM over any of these NP extensions. Nevertheless, even if current data do not point towards the evidence of any specific NP contribution, it still allows us to set bounds to the parameter space of each of these scenarios.

\begin{table}[!t!]
    \centering
    \begin{tabular}{|c|c|c|c|c|c|}
    \hline
    &&&&&\\[-1em]
    NP scenario & $g_{V_R}^{N,e}$ & $g_{T_L}^{N,e}$ & $g_{S_L}^{N,e}$ & $g_{S_R}^{N,e}$ & all\\[2pt]
    \hline
    $\Delta IC$ & 0.6 & 0.1 & 2.2 & 2.2 & 3.1 \\
    \hline
    \end{tabular}
    \caption{Computed values for the goodness-of-fit evaluator $\Delta IC\equiv IC_{\rm NP} - IC_{\rm SM}$ for NP scenarios where the HSN is produced in association with an electron. A positive (negative) value for $\Delta IC$ implies a preference for the SM (NP) scenario.}
    \label{tab:DIC_e}
\end{table}

Going in the details of each case, we observe a similar situation in the vectorial and tensorial scenarios. Indeed, in both cases a non vanishing value for the WC is found in the 68.27\% HPDI, even if a compatibility with zero is observed in the 95.45\% one for both scenarios. While this cannot be interpreted as a statistically significant evidence for any of those WCs, this pattern is however reproduced also by the $\Delta IC$ relative to these model, which still prefer the SM solution, but nevertheless at a lower degree when compared to the scalar cases. Numerically, the higher bounds set by the 95.45\% HPDI correspond to $|g_{V_R}^{N,e}| \leq 0.28$ and $|g_{T_L}^{N,e}|\leq 0.07$.

\begin{figure*}[!t!]
\centering
\includegraphics[width = 0.55\textwidth]{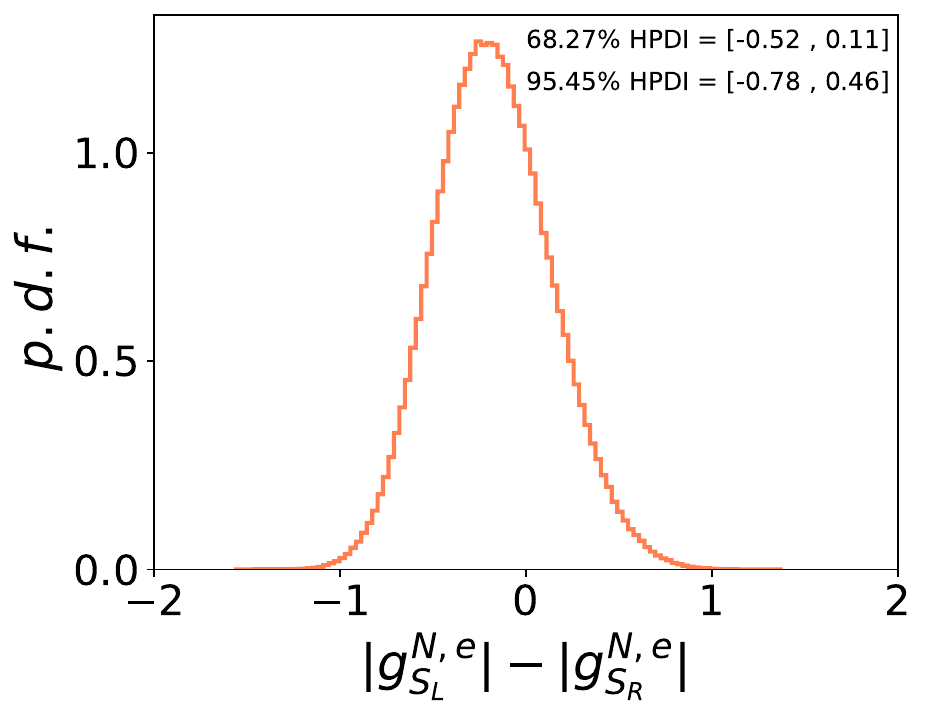}
\caption{Normalised posterior distribution function for the difference of the scalar WCs absolute values, where the HSN is produced in association with an electron. All NP WCs are allowed to be non-vanishing at the same time. 68.27\% and 95.45\% HPDI are reported as well, see the text for further details.}
\label{fig:4WC_e}
\end{figure*}

Concerning the scalar scenarios involving $|g_{S_L}^{N,e}|$ or $|g_{S_R}^{N,e}|$, the obtained results for the two cases are the same due to the structure of $H_S^N$ and $H_P^N$ shown at Eq.~\eqref{eq:HHAs}. Indeed, the only difference among the left-handed coefficient and the right-handed one is the overall sign in the amplitude; however, this difference is washed out once the absolute values of these amplitudes are inserted in the $G_m^{lk}$ angular coefficients at Eqs.~\eqref{eq:G000}-\eqref{eq:G222} and, in the absence of vector/tensor NP WCs, terms proportional to real or imaginary parts of these amplitudes vanish. Numerically, the 95.45\% HPDI for $|g_{S_{L(R)}}^{N,e}|$ set higher bounds corresponding to $|g_{S_{L(R)}}^{N,e}| \leq 0.72$. 

As a following step, we performed a more generic study where all 4 NP WCs are allowed at the same time. The fit results concerning $|g_{V_R}^{N,e}|$ and $|g_{T_L}^{N,e}|$ are identical to the ones obtained in the single WC scenarios reported in Fig.~\ref{fig:1WC_e}, still finding higher bounds set by the 95.45\% HPDI corresponding to $|g_{V_R}^{N,e}| \leq 0.28$ and $|g_{T_L}^{N,e}|\leq 0.07$. No statistically relevant correlation has been found among the two WCs. 

On the other hand, the situation for the scalar WCs is dramatically different. As already observed before, the left-handed and right-handed WCs enter in the HAs always with opposite sign; this fact, together with the absence of deviation from SM predictions for the fitted observables, induces a flat direction among the two WCs. Given this premises, once both WCs are allowed in the same fit it is more interesting to study the \emph{p.d.f.} of their relative difference (rather than the individual ones), which we report in Fig.~\ref{fig:4WC_e}. As expected, this quantity is found to be compatible with 0, with the 95.45\% HPDI bounds reading $-0.78\leq |g_{S_L}^{N,e}|-|g_{S_R}^{N,e}| \leq 0.46 $.

The $\Delta IC$ can be found in Tab.~\ref{tab:DIC_e}, pointing again to a preference towards the SM hypothesis. A larger value w.r.t. the ones obtained for the single WC scenarios is due to the increase of the number of model parameters without any particular improvement in the description of data, already reproduced in the SM in a satisfactory way. 

We conclude observing that, due to the absence of strong evidences of deviations in the measured values for $\hat J_{7,e}$ (see discussion at the end of Sec.~\ref{sec:Obs}), no bound can be set on any of the WCs phases $\phi_i$, with all posteriors being found flat over  the whole range.


\subsection{New Physics in the muon channel}\label{sec:NP_mu}

\begin{figure*}[!t!]
\centering
\includegraphics[width = 0.45\textwidth]{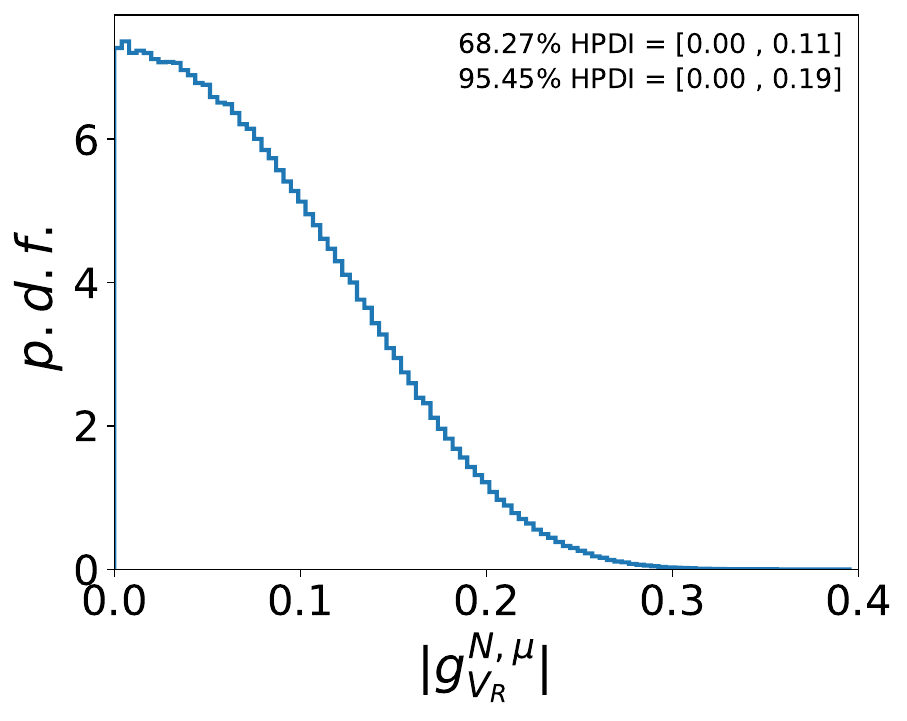}\hspace{1.4em}
\includegraphics[width = 0.47\textwidth]{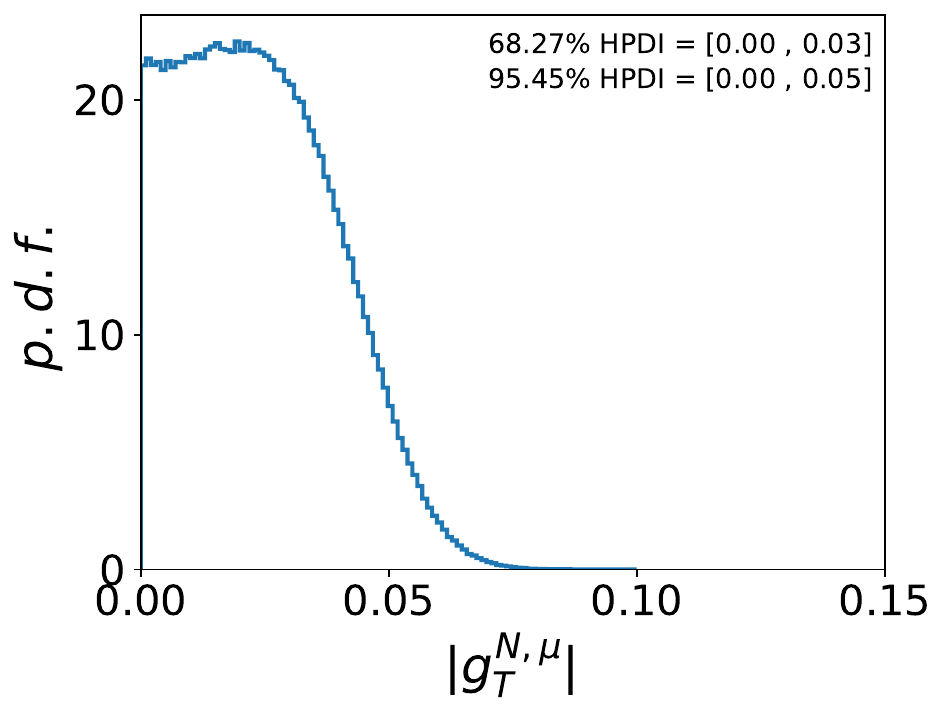}\\
\includegraphics[width = 0.45\textwidth]{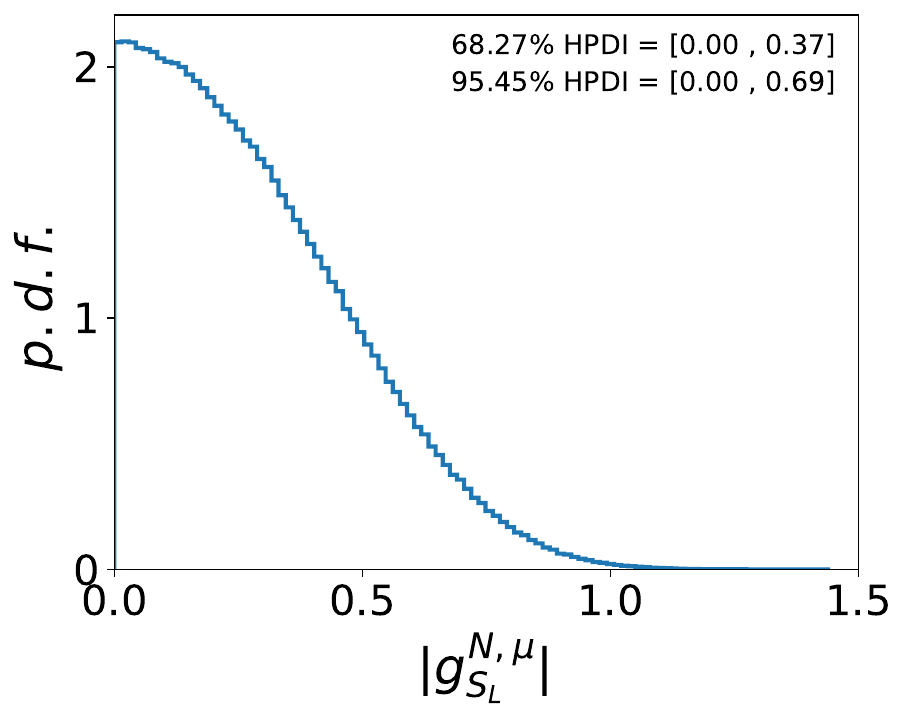}\hspace{2em}
\includegraphics[width = 0.45\textwidth]{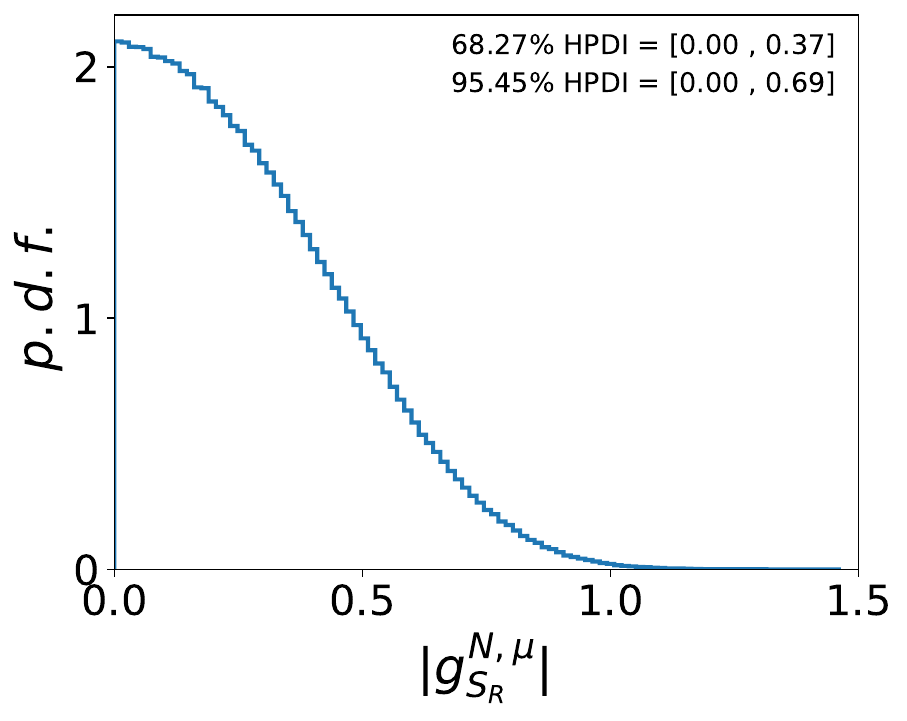}
\caption{Similar to Fig.~\ref{fig:1WC_e}, but for the cases in which the HSN is produced in association with a muon.}
\label{fig:1WC_mu}
\end{figure*}

We move now to the discussion of SM extensions where the HSN is produced in association with muons. Following the same approach of the electron case, we first investigate NP scenarios where we allow for one NP WC at the time, before investigating the scenario where all 4 NP WCs are allowed at the same time. Once again, in the single non-vanishing WC scenarios we will focus only on the coupling absolute values, while also their phases will be allowed to be non-vanishing in the more general case.

\begin{table}
    \centering
    \begin{tabular}{|c|c|c|c|c|c|}
    \hline
    &&&&&\\[-1em]
    NP scenario & $g_{V_R}^{N,\mu}$ & $g_{T_L}^{N,\mu}$ & $g_{S_L}^{N,\mu}$ & $g_{S_R}^{N,\mu}$ & all\\[2pt]
    \hline
    $\Delta IC$ & 2.0 & 1.2 & 2.4 & 2.4 & 6.5 \\
    \hline
    \end{tabular}
    \caption{Computed values for the goodness-of-fit evaluator $\Delta IC\equiv IC_{\rm NP} - IC_{\rm SM}$ for NP scenarios where the HSN is produced in association with a muon. A positive (negative) value for $\Delta IC$ implies a preference for the SM (NP) scenario.}
    \label{tab:DIC_mu}
\end{table}

The \emph{p.d.f.} for the scenarios where we allow only for one WC at the time are shown in Fig.~\ref{fig:1WC_mu}, where once again the marginalized posterior for the absolute values of the 4 WCs are reported. Similarly to the electron case, the posteriors for the HSN mass are flat in each scenario, and no evidence for NP coefficients was found in the muon case as well. This is reflected also in the values for the $\Delta IC$, which for these cases are reported in Tab.~\ref{tab:DIC_mu} and point again to a preference of the SM hypothesis.

The quantitative results for the muon case are mostly similar to the ones obtained for the electron scenario, as a consequence to an analogous measured pattern in the two lepton channels by the Belle collaboration. Similarly to the previous case, the 95.45\% HPDI induce bounds on the vector and tensor case corresponding to $|g_{V_R}^{N,\mu}| \leq 0.19$ and $|g_{T_L}^{N,\mu}|\leq 0.05$; however, no evidence for these WCs is found in the 68.27\% HPDI, as reflected from larger values of the corresponding $\Delta IC$. More similar is the case of scalar WCs, where the 95.45\% HPDI bounds read $|g_{S_{L(R)}}^{N,\mu}| \leq 0.69$. 

\begin{figure*}[!t!]
\centering
\includegraphics[width = 0.47\textwidth]{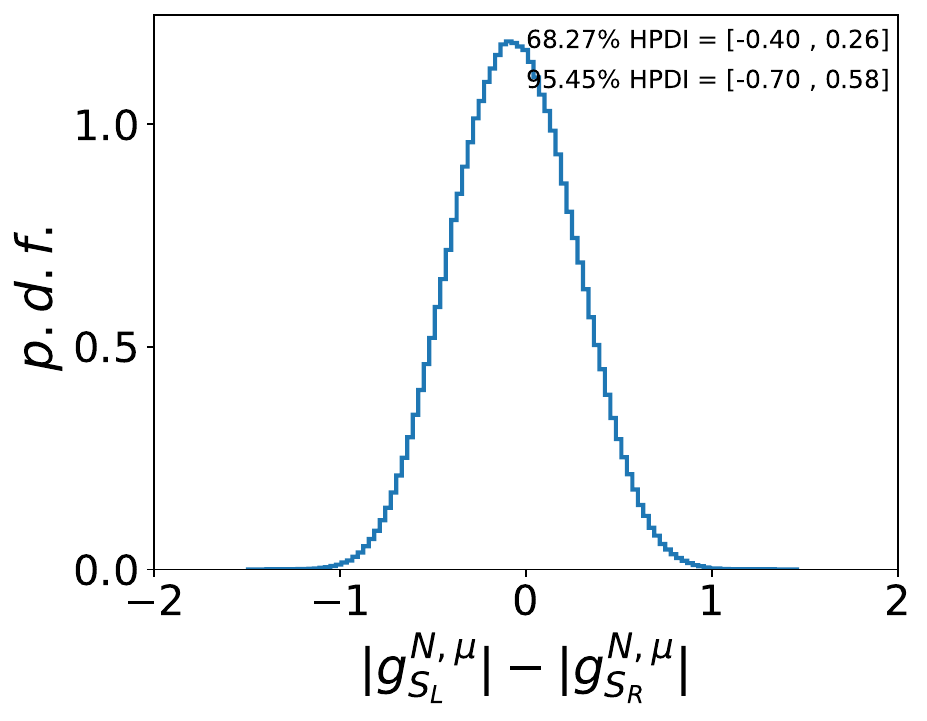}
\caption{Similar to Fig.~\ref{fig:4WC_e}, but for the case in which the HSN is produced in association with a muon.}
\label{fig:4WC_mu}
\end{figure*}

As the last step of our analyses, we studied the NP scenario where all 4 operators with HSN produced together with a muon are allowed to have non-vanishing couplings. Also for this scenario, the results are qualitatively analogous to the ones obtained in the electron case, with no sensitive difference observed for the vector and tensor case when compared to the single WC scenarios. Concerning the scalar WCs, the same flat direction can be found for $|g_{S_L}^{N,\mu}|$ and $|g_{S_R}^{N,\mu}|$, with the obtained \emph{p.d.f.} for their difference given in Fig.~\ref{fig:4WC_mu}. Also this quantity is found to be compatible with 0, with the 95.45\% HPDI bounds reading $-0.69\leq |g_{S_L}^{N,\mu}|-|g_{S_R}^{N,\mu}| \leq 0.59$.

To conclude, the $\Delta IC$ given in Tab.~\ref{tab:DIC_mu}, pointing again to a strong preference towards the SM hypothesis. Moreover, the accordance among the SM prediction for $\hat J_{7,\mu}$ and its measured values implies that again no bound can be set on any of the WCs phases $\phi_i$, with all posteriors being found flat in the whole allowed range.


\section{Comparison with current flavour bounds}\label{sec:comparison}

Now that we have extracted the bounds imposed from the angular $B\to D^*\ell\nu$ data on the HSN effective operators of Eq.~\eqref{eq:Heff_Jp}, it is interesting to study how such bounds fare when confronted with current model-independent limits already present in the literature. In this regard, the main constraints come from the $B\to K^+\nu\bar\nu$ decays. Indeed the SU(2) invariant operators $\epsilon_{a b} (\bar Q_2^a b_R) ( \bar L^b  N_R)$ and $\epsilon_{a b} (\bar Q_2^a \sigma_{\mu\nu} b_R)  (\bar L^b \sigma^{\nu\mu} N_R)$ link the last two terms in Eq.~\eqref{eq:Heff_Jp} to
\begin{equation}
\mathcal{O}_{S_R}^{N,\nu_\ell}=g_{S_R}^{N,\nu_\ell} (\sbar_L b_R) (\nubar_{\ell,L} N_R) \,, \qquad\qquad
\mathcal{O}_{T}^{N,\nu_\ell}=g_{T}^{N,\nu_\ell} (\sbar_L \sigma_{\mu\nu} b_R) (\nubar_{\ell,L} \sigma^{\mu\nu} N_R) \,,
\end{equation}
whose couplings $g_{S_R}^{N,\nu_\ell}$ and $g_{T}^{N,\nu_\ell}$ are the same as $g_{S_R}^{N,\ell}$ and $g_{T}^{N,\ell}$, up to a $V_{cs}$ factor and one-loop or higher-order corrections. The emergence of these operators induces a shift in the $B\to K^+\nu\bar\nu$ decay rate, which can be parameterized (neglecting effects from final state masses) as~\cite{Robinson:2018gza,Kamenik:2009kc,Kamenik:2011vy}
\begin{align}
	\frac{d\Gamma_{B\to K\nu\bar\nu}}{dz}\Big/ \frac{d \Gamma_{B\to K\nu\bar\nu}}{dz}\biggr|_{\rm SM}&=1+ z \frac{32\pi^2}{3 \alpha^2}\bigg|\frac{V_{cb}}{C_{\nu\nu}^{\rm SM} V_{tb} V_{ts}^*}\bigg|^2  \bigg[\frac{3}{8} \frac{\big(g_{S_R}^{N,\nu_\ell}\big)^2}{(1-z)^2} \frac{f_0^2}{f_+^2} + \big(g_{T}^{N,\nu_\ell}\big)^2 \frac{f_T^2}{f_+^2}\bigg] \nonumber 
	\\
	&	\simeq 1+ 5 \times 10^{4}\, z  \bigg[\frac{3}{8}\frac{\big(g_{S_R}^{N,\nu_\ell}\big)^2}{(1-z)^2}  \frac{f_0^2}{f_+^2} + \big(g_{T}^{N,\nu_\ell}\big)^2 \frac{f_T^2}{f_+^2} \bigg], \label{eq:BKnunu}
\end{align}
where $z\equiv q^2/m_B^2$, $f_0(q^2)$, $f_+(q^2)$ and $f_T(q^2)$ are the three form factors entering the $B\to K^+\nu\bar\nu$ decay, and $C_{\nu\nu}^{\rm SM}\simeq -6.35$ is the only WC present in the SM, which implies the prediction $BR(B\to K\nu\bar\nu)|_{\rm SM} \simeq 4 \times 10^{-6}$~\cite{Buras:2014fpa}. Even considering the recent measurement by Belle II, which observed this decay for the first time and found it above the SM expectations, namely at $BR(B\to K\nu\bar\nu) = (2.3 \pm 0.7) \times 10^{-5}$~\cite{Belle-II:2023esi}, Eq.~\eqref{eq:BKnunu} imposes strong bounds on the $g_{S_R}^{N,\ell}$ and $g_{T}^{N,\ell}$ WCs. Indeed, in order to comply with the measured value for $BR(B\to K\nu\bar\nu)$, these coefficients have to be strongly suppressed, at the level of $\mathcal{O}(10^{-2})$ (for a detailed analyses of HSN contributions to this channel, see Refs.~\cite{Felkl:2023ayn,Rosauro-Alcaraz:2024mvx}). 

In conclusion, concerning the tensor WCs and, even more strongly, the right-handed scalar ones, the bounds obtained form the angular analysis of $B\to D^*\ell\bar\nu$ decays in Sec.~\ref{sec:bounds} are not competitive to the ones that can be extracted from an analysis to $B\to K^+\nu\bar\nu$ decays. On the other hand, no competitive flavour bounds are currently available in the literature for the right-handed vector and the left-handed scalar WCs, to the best of our knowledge.


\section{Conclusions}\label{sec:conclusions}

In this paper, we have studied the possible bounds that can be imposed from the recent measurements~\cite{Belle:2023xgj} of the full angular distributions of $B\to D^*\ell\bar\nu$ decays, with $\ell=e,\mu$, to extensions of the SM which include HSN. These studies have been carried out in a model-independent approach, introducing an EFT which includes the lowest dimension four-fermion operators describing the emissions of HSN associated with a light charged lepton $\ell$, together with a $b$ quark and a $c$ quark~\cite{Robinson:2018gza}. In order to be kinematically allowed, the new particle mass should not be heavier than the kinematical threshold $m_N^{\rm max}= (m_B-m_{D^*})^2 \simeq 3$ GeV; however, due to experimental assumptions employed in Ref.~\cite{Belle:2023xgj}, we considered here only the possibility of a HSN with a mass up to 65 MeV, so that we could not study the hint around $M_\mathrm{miss}^2\sim (354\,\mbox{MeV})^2$ first observed here in the bump hunt of Sec.~\ref{sec:Exp}, for which a dedicated Belle II analysis would be required. Nevertheless, the measured $B\to D^*\ell\bar\nu$ angular distribution would still emerge as the incoherent sum of the SM process $B\to D^*\ell\bar\nu_\ell$ and the NP one $B\to D^*\ell\bar N$. We therefore computed the HAs relative to the latter process, and inspected how the inclusion of those alters the predictions of the observed angular observables. Hence, a series of fits to the WCs mediating the NP currents was performed, in order to extract bounds on these couplings from this recently available data.

In order to maintain an agnostic approach, we assumed no LFU violation in the NP contributions to this channel and hence fitted separately for scenarios where the HSN is produced in association with an electron or a muon. More in details, for each of the two different charged lepton scenarios separately, we first performed fits where we allowed only one of the associated HSN WCs to be non-vanishing, before performing a more general fit where all 4 couplings are allowed at the same time. We therefore investigated a total of 10 different scenarios.

The results obtained for the electron and the muon channels are not qualitatively different, mainly due to the fact that the measured angular distributions do not exhibit sensitive deviations from their SM predictions. In particular, no statistically relevant evidence for any NP WCs was obtained in any of the 10 inspected scenarios. Similarly, no preference of a massive HSN was observed, with the mass posterior \emph{p.d.f.} resulting flat over the whole scanned region in all cases. Quantitatively, the 95.45\% HPDI ranges obtained for the WCs in the single coefficient scenarios all include their vanishing values. The strongest bounds were observed for the tensorial WCs at the order of $\sim$ 0.05, while the looser bounds were found for the scalar ones at the order of $\sim$ 0.7; the vectorial WCs obtained bounds sit in the middle, around $\sim$ 0.25. As stated above, these limits are not strongly different between the electron scenario and the muon one.

While these bounds are not sensibly altered in the 4 coupling scenarios for the vector and tensor operator, the situation changes for the two scalar ones. Indeed, these couplings always enter the HAs with a relative opposite sign; therefore, when they are allowed to be non-vanishing at the same time, the lack of a requirement for NP effects translates in a fine-tuning of the two coefficients, whose difference is required to vanish.

Finally, we confronted the bounds here derived with the ones previously obtained in the literature. The main competing constraint comes from another flavour observable, namely the $BR(B\to K\nu\bar\nu)$ recently observed by the Belle II collaboration~\cite{Belle-II:2023esi}. Indeed, the $SU(2)_L$ invariance of the operators mediating the $b\to c\ell\bar N$ induces also ones entering $b\to s\nu\bar N$, with the WCs involved in this neutral current strongly correlated to the ones mediating the charged one. Due to the relative suppression of the former current w.r.t. the latter, the measurement of the $B\to K\nu\bar\nu$ decay induces very stringent bounds on these WCs, even if observed with a rate a few times higher than what predicted by the SM. 

We therefore obtained that the bounds on the vector and the right-handed scalar WCs coming from the $B\to K\nu\bar\nu$ decay and equal to $\mathcal{O}(10^{-2})$ are more competitive than the ones we obtained from $B\to D^*\ell\bar\nu$ decays. On the other hand, the bounds we derived here for the vector and the left-handed scalar operators, in the HSN mass range that we inspected, are the most stringent to date.


\acknowledgments The authors wish to thank José Zurita and Marco Ardu for useful discussions. This research was supported the BMBF grant 05H21VKKBA, \emph{Theoretische Studien f\"ur Belle II 
und LHCb}. MF also enjoyed support from the Generalitat Valenciana (Grant PROMETEO/2021/071) and by MCIN/AEI/10.13039/501100011033 (Grant No. PID2020-114473GB-I00). MP is supported by the German Research Foundation (DFG) Emmy-Noether Grant No. 526218088.


\appendix


\section{Form Factors definitions and determinations}\label{app:FF}

The hadronic matrix elements between a $B$ and a $D^*$ vector can be parameterized, in terms of the 7 form factors $V(q^2)$, $A_0(q^2)$, $A_1(q^2)$, $A_2(q^2)$, $T_1(q^2)$, $T_2(q^2)$ and $T_3(q^2)$, as
\begin{align}
\langle\Dst(k,\varepsilon)|\cbar\gamma_\mu b|\Bbar(p)\rangle =& -i\epsilon_{\mu\nu\alpha\beta}\varepsilon^{*\nu}p^\alpha k^\beta \frac{2V(q^2)}{m_B+m_\Dst} \,, \\
\langle\Dst(k,\varepsilon)|\cbar\gamma_\mu\gamma_5 b|\Bbar(p)\rangle =& \varepsilon_{\mu}^* (m_B+m_\Dst)A_1(q^2) - (p+k)_\mu(\varepsilon^*q)\frac{A_2(q^2)}{m_B+m_\Dst} \nonumber \\
-&q_\mu(\varepsilon^*q)\frac{2m_\Dst}{q^2}\Bigl[ \frac{m_B+m_\Dst}{2m_\Dst} A_1(q^2) - \frac{m_B-m_\Dst}{2m_\Dst} A_2(q^2)-A_0(q^2)\Bigr] \,, \\
\langle\Dst(k,\varepsilon)|\cbar b|\Bbar(p)\rangle =& 0 \,, \\
\langle\Dst(k,\varepsilon)|\cbar\gamma_5 b|\Bbar(p)\rangle =& -(\varepsilon^* q)\frac{2m_\Dst}{m_b+m_c}A_0(q^2) \,, \\
\langle\Dst(k,\varepsilon)|\cbar\sigma_{\mu\nu} b|\Bbar(p)\rangle =& \epsilon_{\mu\nu\alpha\beta} \left[ -\varepsilon^{*\alpha}(p+k)^\beta T_1(q^2) + \varepsilon^{*\alpha}q^\beta \frac{m_B^2-m_\Dst^2}{q^2}[T_1(q^2)-T_2(q^2)] \right. \nonumber \\
& \left. \qquad + (\varepsilon^* q)p^\alpha k^\beta \frac{2}{q^2} \left[ T_1(q^2)-T_2(q^2) - \frac{q^2}{m_B^2 - m_\Dst^2} T_3(q^2) \right] \right] \,, \\
\langle\Dst(k,\varepsilon)|\cbar\sigma_{\mu\nu} \gamma_5 b|\Bbar(p)\rangle =& i \bigg\{ -\left[ \varepsilon_\mu^*(p+k)_\nu - (p+k)_\mu\varepsilon_\nu^* \right] T_1(q^2) \nonumber \\
& + \left[ \varepsilon_\mu^*q_\nu - q_\mu\varepsilon_\nu^* \right] \frac{m_B^2-m_\Dst^2}{q^2}[T_1(q^2)-T_2(q^2)] \nonumber \\
& \left. + \left( \varepsilon^*q \right) \left[ p_\mu k_\nu - k_\mu p_\nu \right] \frac{2}{q^2} \left[ T_1(q^2)-T_2(q^2) - \frac{q^2}{m_B^2 - m_\Dst^2} T_3(q^2) \right] \right\} \,,
\label{eq:FF_Dst}
\end{align}
where $q=p-k$. The convention used for the $\epsilon$ tensor reads $\epsilon_{0123}=1$ (or equivalently $\epsilon^{0123}=-1$). It is useful to introduce the following combinations of form factors:
\begin{align}
A_{12} &= \frac{\left( \mB + \mDs \right)^2 \left (\mB^2 - \mDs^2 - q^2 \right) A_1 - \lambda_B A_2}{16 \mB \mDs^2 \left(\mB + \mDs \right)} \,, \\
T_{23} &= \frac{\left( \mB^2 - \mDs^2 \right) \left (\mB^2 +3 \mDs^2 - q^2 \right) T_2 - \lambda_B T_3}{8 \mB \mDs^2 \left(\mB - \mDs \right)} \,.
\end{align}

In the SM only there are no tensor currents and hence only 4 form factors appear, namely $V(q^2)$, $A_0(q^2)$, $A_1(q^2)$ and $A_2(q^2)$, for which several estimates have been computed in the latest years. Recently, three different determinations of these form factors have been obtained employing Lattice QCD techniques beyond zero recoil, namely by the FNAL/MILC collaboration~\cite{FermilabLattice:2021cdg}, by the HPQCD collaboration~\cite{Harrison:2023dzh} and by the JLQCD collaboration~\cite{Aoki:2023qpa}. 

The form factor basis employed by the  Lattice groups differs from the one here defined, with the relations among the Lattice form factors $f(q^2)$, $g(q^2)$, $\mathcal{F}_1(q^2)$ and $\mathcal{F}_2(q^2)$ and those defined above being
\begin{align}
V(q^2) &= \frac{m_B+m_{D^*}}{2} g(q^2) \,, \qquad\qquad A_0(q^2) = \frac{1}{2} \mathcal{F}_2(q^2) \,, \nonumber \\
A_1(q^2) &= \frac1{m_B+m_{D^*}} f(q^2) \,, \qquad\qquad A_{12}(q^2) = \frac1{8m_Bm_{D^*}} \mathcal{F}_1(q^2)\,.
\end{align}

Currently, the three Lattice determination are not perfectly compatible among themselves, particularly concerning the actual slopes of the $\mathcal{F}_1(q^2)$ and $\mathcal{F}_2(q^2)$ form factors; moreover, when confronted with presently available differential distribution rates measured by Belle~\cite{Belle:2023bwv} and Belle II~\cite{Belle-II:2023okj}, the form factor determination from JLQCD has been the one found to have a better agreement with data than the FNAL/MILC or HPQCD ones, see e.g. discussion in Refs.~\cite{Fedele:2023ewe,Martinelli:2023fwm,Bordone:2024weh,Martinelli:2024vde}. For this reason, we decided to employ the former lattice result, rather then any of the latter ones (or a combination of all available ones, which is mainly driven by FNAL/MILC and HPQCD results). Nevertheless, we checked for completeness that our findings in Sec.~\ref{sec:bounds} do not qualitatively change when employing a different form factor choice. The reason behind this outcome is twofold. On the one hand, the experimental data currently at hand is still not precise enough to be strongly sensitive to the ``theory error'' which could be associated to the discrepancies among the several form factors determinations, given the level of precision we set on our bounds; this implies a reduced power in its capability to distinguish the different approaches. On the other hand, the form factor parameters enter in the fit essentially in the same way as the NP contributions do, i.e. they are not fixed quantities but rather they are allowed to vary according to their prior (which is a correlated multidimensional gaussian \emph{p.d.f.}, in their case); this means, ultimately, that the posterior for the form factor parameters found by the fits in the three approaches are qualitatively similar, hence implying a qualitatively similar result for the NP parameters as well. As expected, the posterior \emph{p.d.f.} obtained in the JLQCD approach are more similar to their priors than what observed in the other two cases.

We conclude observing that, regarding the tensor form factors $T_1(q^2)$, $T_2(q^2)$ and $T_3(q^2)$, we employ the results obtained employing Heavy Quark Effective Theory in Ref.~\cite{Bernlochner:2017jka}, due to the present lack of an estimate for such form factors on the Lattice.


\section{The SM Helicity Amplitudes}\label{app:SMHA}

Here we give the expression for the HAs entering the description of $B\to D^*\ell\nu_\ell$ decays in the SM. Neglecting terms proportional to light charged lepton masses $m_e$ and $m_\mu$, those read:
\begin{align}
\label{eq:HHAs_SM}
H^{\nu_L}_{V;0} &= i \frac{2 \mB \mDs }{\sqrt{q^2}}  A_{12} \,,\qquad \qquad H^{\nu_L}_{V,\pm}=\frac{i}{4 \left(\mB + \mDs\right)} \left(\pm \sqrt{\lambda_B} \, V- \left(\mB + \mDs\right)^2 A_1 \right) \,,   \nonumber  \\
H_P^{\nu_L} &\simeq H_S^{\nu_L} \simeq 0\;, \qquad\qquad\quad\quad\ \ H^{\nu_L}_{T,0} =H^{\nu_L}_{T,\pm} = 0 \;.
\end{align}
We stress that $H_P^{\nu_L}$ and $H_S^{\nu_L}$ are, in the SM, the HAs proportional to the light charged lepton masses and to the form factor $A_0(q^2)$.

\bibliographystyle{JHEP}
\bibliography{hepbiblio}
\end{document}